\newcommand{\PaperTitle}{Browsing without Third-Party Cookies:\\What Do You See?}
\newcommand{\RunningHead}{Browsing without Third-Party Cookies: What Do You See?}
\newcommand{\PaperNumber}{212}

\documentclass[sigconf]{acmart}

\usepackage{amsmath}
\usepackage{caption}
\usepackage{subcaption}
\usepackage{algorithm}
\usepackage[noend]{algpseudocode}
\usepackage[group-separator={,},
            group-minimum-digits=4]{siunitx}
\usepackage{svg}
\usepackage{float}
\usepackage{xspace}
\usepackage[normalem]{ulem}
\usepackage{enumitem,kantlipsum}
\usepackage{tabularray}
\SetTblrInner{rowsep=0pt}

\usepackage{url}
\def\UrlBreaks{\do\/\do-}

\AtEndPreamble{\hypersetup{colorlinks,
linkcolor=ACMPurple,
citecolor=ACMPurple,
urlcolor=ACMDarkBlue,
filecolor=ACMDarkBlue}}

\usepackage[compact]{titlesec}
\setitemize{noitemsep,topsep=3pt,parsep=0pt,partopsep=0pt, leftmargin=*}
\setenumerate{noitemsep,topsep=3pt,parsep=0pt,partopsep=0pt, leftmargin=*}

\author{Maxwell Lin}
\orcid{0009-0006-5672-7403}
\affiliation{%
  \institution{Duke University}
  \department{Computer Science}
  \city{Durham}
  \state{North Carolina}
  \country{USA}
}
\email{maxwell.lin@duke.edu}

\author{Shihan Lin}
\orcid{0000-0001-7182-9041}
\affiliation{%
  \institution{Duke University}
  \department{Computer Science}
  \city{Durham}
  \state{North Carolina}
  \country{USA}
}
\email{shihan.lin@duke.edu}

\author{Helen Wu}
\orcid{0009-0005-6213-0083}
\affiliation{%
  \institution{Vanderbilt University}
  \department{Computer Science}
  \city{Nashville}
  \state{Tennessee}
  \country{USA}
}
\email{helen.wu@vanderbilt.edu}

\author{Karen Wang}
\orcid{0009-0007-1906-4345}
\affiliation{%
  \institution{Duke University}
  \department{Computer Science}
  \city{Durham}
  \state{North Carolina}
  \country{USA}
}
\additionalaffiliation{%
  \institution{Duke Kunshan University}
  \department{Applied Mathematics and Computational Sciences}
  \city{Kunshan}
  \state{Jiangsu}
  \country{China}
}
\email{karen.wang@duke.edu}

\author{Xiaowei Yang}
\orcid{0000-0002-0664-3846}
\affiliation{%
  \institution{Duke University}
  \department{Computer Science}
  \city{Durham}
  \state{North Carolina}
  \country{USA}
}
\email{xwy@cs.duke.edu}

\copyrightyear{2024}
\acmYear{2024}
\setcopyright{rightsretained}
\acmConference[IMC '24]{Proceedings of the 2024 ACM Internet Measurement Conference}{November 4--6, 2024}{Madrid, Spain}
\acmBooktitle{Proceedings of the 2024 ACM Internet Measurement Conference (IMC '24), November 4--6, 2024, Madrid, Spain}
\acmDOI{10.1145/3646547.3689014}
\acmISBN{979-8-4007-0592-2/24/11}

\makeatletter
\gdef\@copyrightpermission{
 \begin{minipage}{0.3\columnwidth}
 \href{https://creativecommons.org/licenses/by/4.0/}{\includegraphics[width=0.90\textwidth]{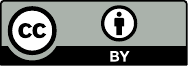}}
 \end{minipage}\hfill
 \begin{minipage}{0.7\columnwidth}
 \href{https://creativecommons.org/licenses/by/4.0/}{This work is licensed under a Creative Commons
Attribution International 4.0 License.}
 \end{minipage}
 \vspace{5pt}
}
\makeatother

\begin{document}
\settopmatter{printfolios=true} 

\begin{abstract}
Third-party web cookies are often used for privacy-invasive behavior
tracking. Partly due to privacy concerns, browser vendors have started to
block \emph{all} third-party cookies in recent years.
To understand the effects of such third-party cookieless browsing, 
we crawled and measured the top 10,000 Tranco websites.
We developed a framework to remove third-party cookies 
and analyze the differences between the appearance of web pages with and without these cookies.
We find that disabling third-party cookies has no substantial
effect on website appearance including layouts, text, and images. 
This validates the industry-wide shift towards cookieless browsing as a way to
protect user privacy without compromising on the user experience.
\end{abstract}



\title[\RunningHead]{\PaperTitle}
\maketitle

\section{Introduction}
\label{sec:introduction}
Websites use web cookies to record stateful information in a browsing session. While some cookies are necessary for websites to work properly (e.g., authentication cookies), the majority of cookies are used for user tracking and advertising~\cite{huCCCCCorrallingCookies2021}, raising concerns about user privacy. In a web measurement study, Englehardt et al. found that an adversary can reconstruct 62-73\% of a typical user’s browsing history using third-party cookies~\cite{englehardtCookiesThatGive2015}. Besides, Papadopoulos et al. found that User IDs are leaked, on average, to 3.5 different domains using cookie synchronization techniques~\cite{papadopoulosCookieSynchronizationEverything2019}.

Because of general privacy concerns, many governments have enacted regulations that require websites to implement cookie notices. These cookie notices allow users to consent to or opt out of the use of unnecessary cookies. For example, the California Consumer Privacy Act (CCPA)~\cite{CaliforniaConsumerPrivacy2018} is a regulation that requires websites to provide a clear opt-out mechanism for their users. In the European Union, the General Data Protection Regulation (GDPR)~\cite{GDPR2016a} and ePrivacy Directive~\cite{Directive2002582002} require websites to obtain specific, informed, and unambiguous user consent before accessing or storing any non-essential user data.  Previous studies have measured the effect of privacy regulations on the cookie landscape.
Degeling et al. discovered that the use of cookie notices increased by 16\% in EU member states after GDPR went into effect ~\cite{degelingWeValueYour2019}. Furthermore, cookie regulations have given rise to Consent Management Platforms (CMPs) which provide ``consent as a service'' solutions to websites. Due to their ease of use, many websites employ CMPs to implement cookie notices and manage the dissemination of user consent to third parties.

Although cookie notices are intended to give users more control over their privacy, 88\% of cookie notices incorporate design
techniques that hinder the ability of users to select
privacy-protective options~\cite{habibOkayWhateverEvaluation2022}. In
a user study, Habib et al. found that 73\% of participants
opted for the most permissive cookie
settings and only 45\% of the participants
chose the settings that genuinely reflected their desired level of
consent~\cite{habibOkayWhateverEvaluation2022}.
  This result indicates
  that cookie notices are an unreliable way for users to choose their
  desired privacy settings.

As a potential solution, \textit{notice-centric} methods have been
proposed to automate user-interaction with cookie notices.  For
example, Khandelwal et al. developed \textit{CookieEnforcer}, a browser extension which can automatically disable all unnecessary cookies with 94\%
accuracy~\cite{khandelwalAutomatedCookieNotice2023}. However, notice-centric methods assume that websites will honor the choices that users make, which may not be the case
in practice.

In a preliminary study (\S~\ref{sec:background}), we found $255$ websites with cookie notices from the top $350$ Tranco domains. Out of these $255$ websites, we found that 100 (39\%) of them do not respect the user's choice and
continue to use tracking cookies even after the user opts out in the
provided cookie notice. Even websites that employ
CMPs violate cookie regulations. We crawled $118$ websites that use
\textit{OneTrust}, the most popular CMP, and we found that $58\%$ use
tracking cookies even when they are disabled via the OneTrust
API.
Lastly, by design, notice-centric methods cannot work if the website does not provide a
cookie notice in the first place. In a large scale study of over
17,000 websites, Kampanos et al. found that less than $50\%$ of
websites show a cookie notice to the
user~\cite{kampanosAcceptAllLandscape2021}.
Therefore, notice-centric cookie enforcement will be unsuccessful on a majority
of websites.

Due to limitations in the notice-centric approach, an ideal approach may be creating a policy framework that enables a user to decide whether a cookie should be enabled without depending on the the presence of a cookie notice. A privacy-conscious user may enable only strictly necessary cookies. However, this approach requires a cookie classifier, and as we explain in \S~\ref{sec:background}, categorizing cookies by function is difficult.

Partly due to the abovementioned challenges of disabling privacy-invasive cookies, 
a recent trend across web browsers has been to disable \textit{all} third-party cookies. 
This approach resolves the privacy concerns caused by third-party cookies, and it is easy to implement and requires little maintenance.   
Google Chrome, the most popular web browser, is following this approach and
will deprecate all third-party cookies in early
2025~\cite{PreparePhasingOut}. Other popular browsers that follow this approach include Safari~\cite{FullThirdPartyCookie2020}, which started blocking all third-party cookies by default since 2020, and Firefox~\cite{LatestFirefoxRolls2018}, which started blocking cross-site tracking cookies using the Disconnect~\cite{Disconnect2024} blocklist since 2018. In this paper, we refer to this approach as \textit{cookieless browsing}, a term which is already used in industry~\cite{teamGetReadyCookieless2024, WIREDBrandLab2024}.

While cookieless browsing certainly limits websites from tracking
users, it remains unclear whether disabling all third-party cookies
impacts the user experience of websites. This ambiguity motivates our study. In this paper, we design and implement a framework to rigorously assess the influence of cookieless browsing on key website operations, such as layout rendering and text/image content. We seek to answer the question: \textit{How does third-party cookieless browsing affect the way websites are displayed to users?} 

Overall, our contributions are summarized as follows:
\begin{itemize}
    \item We design a framework to measure how removing third-party cookies impacts website rendering. To our knowledge, this is the first large-scale study analyzing this behavior. We implement this framework using a Selenium-based web crawler and use it to measure the top 10,000 Tranco~\cite{LePochat2019} domains.
    \item By analyzing extracted features (screenshots, text, images, and links) across different crawl groups, we find that the absence of third-party cookies does not substantially affect the rendering of websites. More than 90\% of the domains showed minimal changes in website rendering, suggesting that cookieless browsing may not degrade user experience.
    \item We release our implementation and results for reproducibility and further analysis.\footnote{See \url{https://github.com/maxwellmlin/IMC2024-CookielessBrowsing}.}
\end{itemize}

\begin{table}[t]
    \centering
    \begin{tblr}{l|c|c|c}
         \textbf{Category} & \textbf{Total} & \textbf{Mean} &\textbf{Median} \\
         \hline
         Strictly Necessary & 940 & 1.83 & 1 \\
         \hline
         Performance & 905 & 1.76 & 0 \\
         \hline
         Functionality & 103  &0.20 & 0 \\
         \hline
         Tracking & 1507 & 2.94 & 0 \\
         \hline
         Unclassified & 14619 & 28.50 & 18
    \end{tblr}
    \caption{The total, mean, and median of cookie counts in each category identified by \textit{Cookie-Script}~\cite{CookieScript} across 255 websites. Note that the vast majority of cookies are unclassified.}
    \label{tab:cookie_script_category}
\end{table}

\section{Background \& Related Work}
\label{sec:background}

\emph{ICC UK cookie categories.}
To facilitate user privacy protection, the International Chamber of Commerce UK (ICC UK) defines four types of cookies~\cite{ICCUKCookie2012}:
\begin{enumerate}
    \item \textbf{Strictly Necessary Cookies:} Enable users to move around the website and use requested features, such as accessing secure areas of the website or adding items to a shopping cart.
    \item \textbf{Performance Cookies:} Collect anonymized information about how visitors use a website (e.g., popular pages, error logs).
    \item \textbf{Functionality Cookies:} Remember choices that users make (such as username, language, or region) to provide personalized features. 
    \item \textbf{Tracking Cookies:} Collect information about users' browsing habits to deliver relevant advertisements.
\end{enumerate}

\begin{figure}[t]
    \centering
    \begin{subfigure}[t]{0.49\linewidth}
        \centering
        \includegraphics[width=\linewidth]{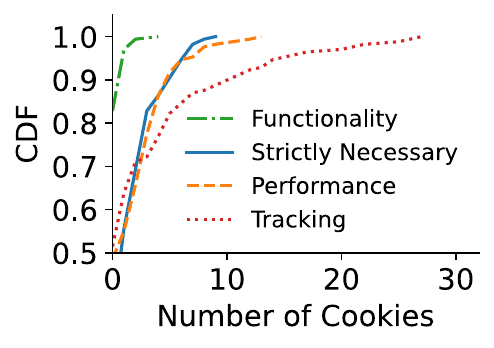}
        \caption{After accepting cookies\label{fig:accept_cookies}}
    \end{subfigure}
    \hfill
    \begin{subfigure}[t]{0.49\linewidth}
        \centering
        \includegraphics[width=\linewidth]{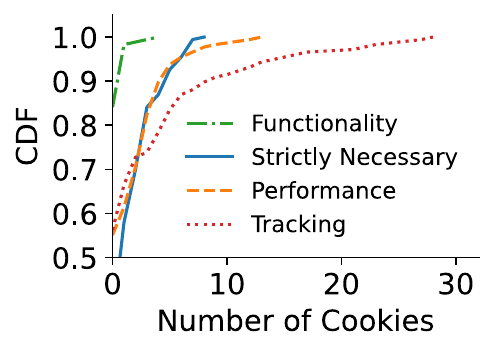}
        \caption{After rejecting cookies\label{fig:reject_cookies}}
    \end{subfigure}
    \Description{2 CDF plots showing the number of cookies collected after clicking either the accept or reject button in a cookie notice. The 2 distributions are similar, suggesting that most websites do not respect user choices in cookie notices.}
    \caption{\label{fig:cookie_script_boxplot}Distribution of cookie counts after cookie notice interaction. Note that the curves in Figure~\ref{fig:accept_cookies} and Figure~\ref{fig:reject_cookies} are very similar, suggesting that most websites do not respect user choices in cookie notices.}
\end{figure}

\emph{Automated cookie classification.}
Many cookie databases attempt to map cookies to their ICC UK category. For example, \textit{Cookie-Script}~\cite{CookieScript} will categorize all present cookies in a given website as either one of the four ICC UK categories or the \textit{unclassified} category if no database entry is found. In a preliminary study, we crawled $255$ websites with cookie notices from the top 350 Tranco domains and categorized collected cookies using \textit{Cookie-Script}. 
As Table~\ref{tab:cookie_script_category} shows, the vast majority of cookies cannot be classified. There are roughly ten times more unclassified cookies than any of the four categorized cookie types, and \textit{Cookie-Script} cannot identify the functionality cookies in most websites.
Due to the ever-changing nature of the web, cookie databases will never be fully comprehensive.

A potential solution is to train a machine learning classifier. Hu et al.~\cite{huCCCCCorrallingCookies2021} introduced \textit{CookieMonster}, a machine learning model capable of categorizing cookies with 94\% accuracy. Bollinger et al.~\cite{bollingerAutomatingCookieConsent2022} developed \textit{CookieBlock}, a machine-learning classifier and browser extension which automatically blocks privacy-invasive cookies with 90\% accuracy. Unfortunately, machine-learning models are never completely accurate, and such methods may still be unsatisfactory to a particularly privacy-conscious user.

\emph{Measurement of cookie usage.} Researchers have launched measurements on the web to investigate the privacy leakage caused by third-party cookies~\cite{englehardtCookiesThatGive2015,papadopoulosCookieSynchronizationEverything2019,PAM15-TrackAdvisor}, the cookie compliance with GDPR~\cite{fouadComplianceCookiePurposes2020,degelingWeValueYour2019,bollingerAutomatingCookieConsent2022}, and the characteristics of cookies in the wild~\cite{CCS16-OnlineTracking,rasaiiExploringCookieverseMultiPerspective2023,cahnEmpiricalStudyWeb2016,Security16-WebTracking}. In our preliminary study, we investigated the potential privacy violation caused by websites' disrespect to user choices in cookie notices. Specifically, we adopted \emph{BannerClick}~\cite{rasaiiExploringCookieverseMultiPerspective2023} to automatically click the accept and reject buttons in cookie notices. For each website, we counted the number of cookies injected after \textit{BannerClick} interaction and categorized the collected cookies using \textit{Cookie-Script}. Figure~\ref{fig:cookie_script_boxplot} shows the results. We found that 100 (39\%) of 255 websites still send at least one tracking cookie after a user rejects the cookies. Besides, the curves in Figure~\ref{fig:accept_cookies} and Figure~\ref{fig:reject_cookies} are very similar, suggesting that most websites do not respect user choices in cookie notices.

\section{Web Crawler Architecture}
\label{sec:crawler}

The above observations motivated us to pursue third-party cookieless browsing as a better technique to protect user privacy. However, this approach raises a question of whether third-party cookieless browsing affects the way websites are displayed to users. To answer this question, we developed a web crawler that measures how website rendering is affected by removing all third-party cookies. The following sections describe the crawler architecture: \S~\ref{sec:groups} explains how we use \textit{crawl groups} to isolate the effects of third-party cookies, \S~\ref{sec:clickstream-generation}-\ref{clickstream-traversal} describes how we use \textit{clickstreams} to explore the inner pages of a website, and \S~\ref{sec:features} describes the \textit{extracted features} our crawler collects for later analysis.

Our crawler uses Selenium~\cite{Selenium} to drive a headless Firefox instance. We deploy our crawler on the top 10,000 domains of the Tranco~\cite{LePochat2019} list generated on Feb. 18th, 2024.\footnote{Available at \url{https://tranco-list.eu/list/KJ2GW}.} The total running time across all jobs was $241$ days; the median time to crawl one domain was $33$ minutes. If a domain took more than $60$ minutes, we terminated the process and moved on to the next domain.

\subsection{Domain-to-URL Resolution}

\begingroup
\def\UrlBreaks{\do\/\do-\do\.}
We first resolve the apex domain obtained from Tranco to an accessible URL. An apex domain is a two-level domain~\cite{hoffmanDNSTerminology2019}, e.g., \texttt{example.com}. To resolve the apex domain, we alternate between the \texttt{http} and \texttt{https} protocols in addition to optionally prepending the \texttt{www} subdomain. For example, to resolve the \url{example.com} domain, we attempt navigation to \url{https://example.com}, \url{https://www.example.com}, \url{http://example.com}, and \url{http://www.example.com}. Domain-to-URL resolution is successful if we can retrieve a web page with clickable elements (see \S~\ref{sec:clickstream-generation}).
\endgroup

\subsection{Crawl Groups}\label{sec:groups}
To measure the effects of third-party cookies on website appearance, we crawl each domain three separate times \textit{sequentially} corresponding to the following groups:
\begin{enumerate}
    \item \textbf{Baseline:} We generate and save a length-$5$ clickstream with all cookies enabled using the method shown in \S~\ref{sec:clickstream-generation}.
    \item \textbf{Control:} We traverse the generated clickstream under the same conditions as the \textit{baseline} group. See \S~\ref{clickstream-traversal}.
    \item \textbf{Experimental:} We traverse the generated clickstream with all third-party cookies disabled. See \S~\ref{clickstream-traversal}.
\end{enumerate}

A unique Firefox profile is created for each group to isolate stateful information such as cookies between browsing sessions.

The motivation behind these three groups is as follows: if we observe a substantial difference between the baseline and experimental groups, it is possible that the website requires the use of cookies. However, we cannot conclude that the experimental condition caused this change in behavior since the content of some websites can change naturally upon page reload (e.g., social media websites). To take this into account, we crawl each website twice without applying the experimental condition (i.e., the baseline and control groups). If these two groups are similar but the experimental group is substantially different, then we can conclude that the experimental condition (disabling third-party cookies) likely caused the observed change in website appearance.

\begin{figure}[t]
\begin{center}
    \centering
    \includegraphics[width=\columnwidth]{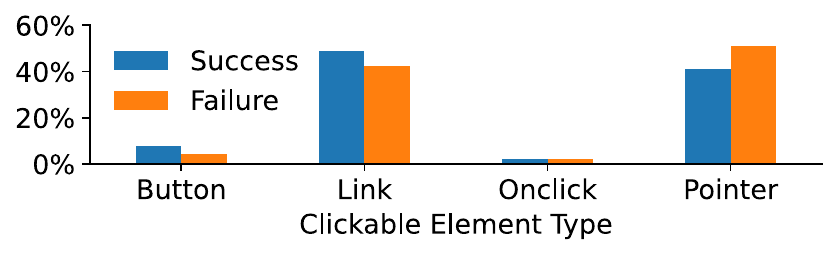}
    \Description{A bar plot comparing the percentage of successful clickable elements and failed clickable elements. Link elements succeed the most often. Pointer elements fail the most often.}
\end{center}
\caption{\label{fig:clickable-elements} Percentage of each element type present within the set of all successful and all failed clickable elements. A clickstream is generated by constructing a sequence of successful clickable elements. When the generated clickstream is later traversed, a clickable element fails when its CSS selector can no longer be resolved.}
\end{figure}

\begin{figure}[t]
\begin{center}
    \centering
    \includegraphics[width=0.9\columnwidth]{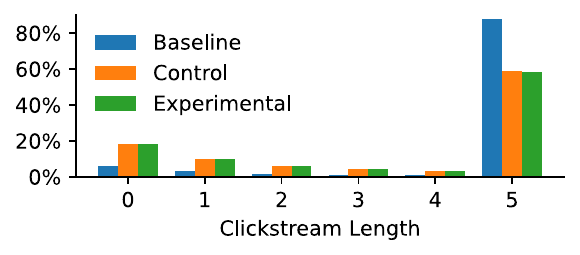}
    \Description{A normed histogram displaying the average clickstream length across the 3 different crawl groups. The baseline group has significantly more successful full-length clickstreams than the control and experimental group. The control and experimental group have comparable clickstream lengths.}
\end{center}
\caption{\label{fig:clickstream_length} Normalized histogram of clickstream length across crawl groups. Clickstream generation occurs in the baseline group while clickstream traversal occurs in the control and experimental groups.}
\end{figure}

\subsection{Clickstream Generation}
\label{sec:clickstream-generation}
A length $k$ clickstream is a sequence of $k$ clickable elements. We define a clickable element as an HTML element that satisfies at least one of the following criteria:
\begin{itemize}[leftmargin=*]
    \item \textbf{Button:} Elements with a \texttt{<button>} tag.
    \item \textbf{Link:} Elements with an \texttt{<a>} tag.
    \item \textbf{Onclick:} Elements with an \texttt{onclick} event attribute.
    \item \textbf{Pointer:} Elements with a "pointer" \texttt{cursor} style.
\end{itemize}

After navigating to the landing page, we sample uniformly at random from the current set of clickable elements until a successful click is made. This successful clickable element is then appended to the current clickstream. This process repeats until the desired clickstream length is generated. If the set of clickable elements is ever empty or we navigate to a different domain, the clickstream prematurely ends. Figure~\ref{fig:clickable-elements} shows the percentage of successful clickable element types encountered during clickstream generation.

Ideally, the clickstream length should be long enough such that the inner pages of a website are thoroughly explored. By analyzing the browsing data of 2.5M users across 760 websites, Lehmann et al. found that users only view $2.36$ pages of a website on average~\cite{lehmannOnlineMultitaskingUser2013a}. Xing et al. examined how users navigate search engine results and found that the majority of users conduct less than $2$ clicks per session~\cite{xingIncorporatingUserPreferences2013a}. Recent research found that the average user clickstream length is 2.5 for e-commerce websites~\cite{ECommerce23-Clickstream}. Thus, a clickstream length of 5 seems appropriate to perform representative inner page exploration. We further justify using a clickstream length of 5 with the following 2 findings:
\begin{enumerate}
    \item Our results in \S~\ref{sec:results} demonstrate that different clickstream lengths do not yield substantially different results. Thus, our conclusions regarding cookieless browsing should not be substantially impacted by the exploration depth.
    \item In Figure~\ref{fig:clickstream_length}, fully traversing a length-5 clickstream has a failure rate of about 40\%. Longer clickstreams become even more difficult to fully traverse as more actions must be successfully completed sequentially.
\end{enumerate}

In our implementation, clickable elements are saved using a unique CSS selector generated by the \texttt{finder}~\cite{medvedevAntonmedvFinder2024} JavaScript library. By using CSS selectors, our crawler is able to traverse the \textit{same} clickstream (and thus, the same subpages) for all 3 crawl groups.
According to an independent benchmark~\cite{fridrichFczbkkCssselectorgeneratorbenchmarkBenchmark2024}, \texttt{finder} was able to consistently generate the shortest selectors compared to $12$ other libraries. Shorter selectors are more reliable across page visits, resulting in more successful clickstream traversals.

\subsection{Clickstream Traversal}
\label{clickstream-traversal}
After navigating to the landing page, we click each element in the clickstream sequentially. Sometimes, the CSS selector originally generated cannot be resolved and we are unable to fully traverse the clickstream. 
This is caused by \textit{dynamic} web pages which have content that changes between page reloads (Figure~\ref{fig:youtube} in Appendix~\ref{app:example} shows an example). Figure~\ref{fig:clickstream_length} presents the normalized histogram of clickstream length across crawl groups. Note that making the clickstream longer reduces the number of successful full-length clickstream traversals. Figure~\ref{fig:clickable-elements} shows the percentage of clickable element types that failed during traversal.

\subsection{Extracted Features}\label{sec:features}
After navigating to the landing page and after each action in a clickstream, we extract the following set of features:
\begin{enumerate}
    \item \textbf{Screenshot:} We take a screenshot of the viewport after scrolling to the top of the page. This ensures that screenshots are consistently aligned, though content below the viewport is not captured in the process. \label{feature:screenshot}
    \item \textbf{Image Shingles:} We also convert the screenshot to an image shingle frequency vector through \textit{Image Shingling}~\cite{andersonSpamscatterCharacterizingInternet2007} as explained later in this section.\label{feature:shingles}
    \item \textbf{Inner Text:} The \texttt{innerText} of the \texttt{<body>} element extracted as a word frequency vector. \label{feature:innerText}
    \item \textbf{Images:} The \texttt{src} attribute of all \texttt{<img>} elements extracted as a frequency vector. \label{feature:images}
    \item \textbf{Links:} The \texttt{href} attribute of all links extracted as a frequency vector. \label{feature:links}
\end{enumerate}

For each domain, we repeatedly generate/traverse new clickstreams until $50$ sets of features are collected for each crawl group.

\textit{Image Shingling}, proposed by Anderson et al., is a method to compare the similarity of images~\cite{andersonSpamscatterCharacterizingInternet2007}. First, we divide each image into fixed sized chunks in memory. Like Anderson et al., we found that an image chunk size of 40x40 pixels was an effective trade-off between granularity and shingling performance. Each chunk is then hashed using MD5 to form an image shingle. Lastly, we extract the frequency vector of image shingles. 
\section{Analysis}
\label{sec:analysis}
In this section, we define and compute two metrics, \emph{BCE (baseline, control, experimental) screenshot difference} and \emph{DiD (difference in distance)}, from the extracted features in \S~\ref{sec:features} to 
compare the websites' visual appearance among the three crawl groups defined in \S~\ref{sec:groups}. Thus, by investigating the statistics of these two metrics, we show the results and conclusions in \S~\ref{sec:results}.

\subsection{Screenshot Comparison}
To compare the Baseline, Control, and Experimental screenshots (Feature~\ref{feature:screenshot}), we define a \emph{BCE screenshot difference} $\Delta \in [0, 1]$ according to Algorithm~\ref{algo:bce_comparison}. This difference excludes the dynamic content in its calculation.

Concretely, we split each crawl group screenshot into 40x40 pixel chunks. Similar to Image Shingling, we found that this chunk size offers good balance between both precision and performance. We filter out all chunks that differ between the baseline and control group to ensure that any naturally occurring differences are excluded. For all remaining chunks, we compute the proportion that differ between the baseline and experimental group. This allows us to only measure differences that occur due to the experimental condition. Note that if baseline and control are identical, we simply return the percent difference between baseline and experimental.

If there are no chunks remaining after the filter (i.e., the baseline and control screenshots are completely different) or the image dimensions are different, then we skip the comparison. This occurred in only 39 (0.39\%) domains spread across the entire Tranco list. Through manual inspection, we find that most of these skipped domains are due to the browser rendering different viewport sizes between crawl groups.

We use Image Shingling rather than other image comparison methods such as keypoint matching~\cite{lourencoSRDSIFTKeypointDetection2012, kelmanKeypointDescriptorsMatching2007}, FSIM~\cite{ToIP11-FSIM}, and SSIM~\cite{ToIP04-SSIM}, because we need to compare the partial regions of screenshots excluding the dynamic content. By dividing the screenshots into chunks, Image Shingling enables such a comparison by comparing a subset of corresponding chunks of the screenshots.

\begin{algorithm}
\caption{BCE screenshot difference}
\label{algo:bce_comparison}
\begin{algorithmic}[1]

\State \(\triangleright\) \text{chunk}() returns a list of chunks given a screenshot from each group (baseline, control, experimental).
\State \textbf{initialize} $\text{chunks} \gets \text{chunk}(\text{screenshots to compare})$
\State \textbf{initialize} $\text{matches} \gets 0$
\State \textbf{initialize} $\text{total} \gets 0$
\State
\State \(\triangleright\) Loop over crawl group chunks in parallel.
\For{\textbf{each} (baseline, control, experimental) in chunks}
    \If{$\text{baseline} == \text{control}$}
        \State $\text{total} \gets \text{total} + 1$
        \If{$\text{baseline} == \text{experimental}$}
            \State $\text{matches} \gets \text{matches} + 1$
        \EndIf
    \EndIf
\EndFor
\State
\State \(\triangleright\) Return a difference $\Delta \in [0, 1]$.
\State \Return $1 - \left(\frac{\text{matches}}{\text{total}}\right)$

\end{algorithmic}
\end{algorithm}

\subsection{Frequency Vector Comparison}
Extracted frequency vectors (Features \ref{feature:shingles}-\ref{feature:links}) are compared using the Jaccard distance formula. If we let $A$ and $B$ be the multisets of two frequency vectors, we define the Jaccard distance as
\begin{align}
    J(A,B):=1-\frac{|A \cap B|}{|A \cup B|} \in [0,1].
\end{align}

If $|A \cup B|=0$, i.e., $A$ and $B$ are both empty, then we define $J(A,B):=0$. In other words, we consider two empty sets to be identical.

We compare the extracted features using a difference in distance approach. The \emph{difference in distance (DiD)} computes the following
\begin{equation} \label{eq:did}
\text{DiD}:=J(B,E) - J(B,C) \in [-1,1]
\end{equation}
where $B$, $C$, $E$ denote a frequency vector from the baseline, control, and experimental groups respectively.

If the DiD is small, then the experimental condition did not have a substantial effect since $J(B,E) \approx J(B,C)$. 

\section{Results}
\label{sec:results}

Out of the $\num[group-separator={,}]{10000}$ input Tranco domains, $\num[group-separator={,}]{1897}$ domains were unable to be resolved to a URL. Through manual inspection, we found that many of these domains (e.g., \texttt{akamai.net}) are used by content delivery networks for DNS redirection. After filtering out other problematic domains (e.g., domains that hang our crawler), we arrive at a final list of $\num[group-separator={,}]{7490}$ successfully crawled domains.

\begin{figure}[t]
\begin{center}
    \centering
    \includegraphics[width=0.85\columnwidth]{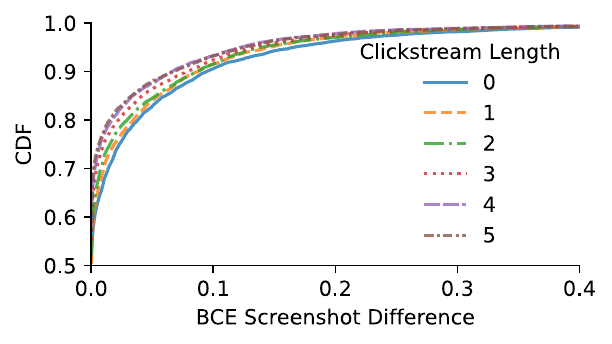}
    \Description{A CDF of BCE screenshot differences, separated by clickstream length. More than 90\% of domains exhibit less than a 10\% screenshot difference when third-party cookies are disabled.}
\end{center}
\caption{\label{fig:bce_diff} CDF of BCE screenshot differences (Algorithm~\ref{algo:bce_comparison}). A clickstream length of 0 means taking a screenshot of the landing page without any clicks.}
\end{figure}

\begin{figure}[t]
\begin{center}
    \centering
    \includegraphics[width=0.9\columnwidth]{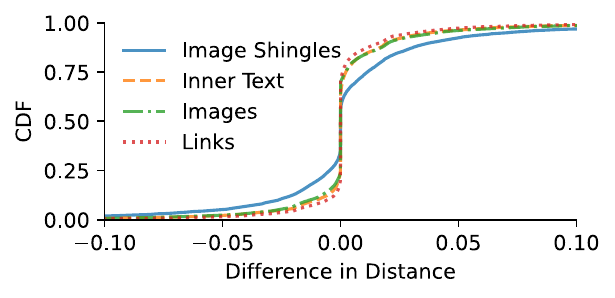}
    \Description{4 CDFs displaying the difference in distance distribution across image shingles, inner text, images, and links. The distribution is concentrated and symmetric at 0.}
\end{center}
\caption{\label{fig:combined-jaccard-did} CDF of frequency vector DiDs (Equation~\ref{eq:did}).}
\end{figure}

\begin{table}[h]
\centering
\begin{tabular}{|l|l|}
\hline
\textbf{Category} & \textbf{Count} \\ \hline
Adult Themes & 638 \\ \hline
Business \& Economy & 1148 \\ \hline
CIPA & 836 \\ \hline
Education & 662 \\ \hline
Entertainment & 1990 \\ \hline
Gambling & 184 \\ \hline
Government \& Politics & 377 \\ \hline
Health & 117 \\ \hline
Internet Communication & 454 \\ \hline
Questionable Content & 73 \\ \hline
Shopping \& Auctions & 560 \\ \hline
Society \& Lifestyle & 477 \\ \hline
Sports & 171 \\ \hline
Technology & 2244 \\ \hline
Travel & 128 \\ \hline
Vehicles & 80 \\ \hline
\end{tabular}
\caption{Categorization counts. Note that a single domain can have more than one category. CIPA in the table refers to Children's Internet Protection Act~\cite{CIPA}. Cloudflare defines Questionable Content as content that is related to hacking, piracy, profanity, or other questionable activities~\cite{CloudflareDocs}.}
\label{tab:category_values}
\end{table}

\begin{figure}[h]
\begin{center}
    \centering
    \includegraphics[width=\columnwidth]{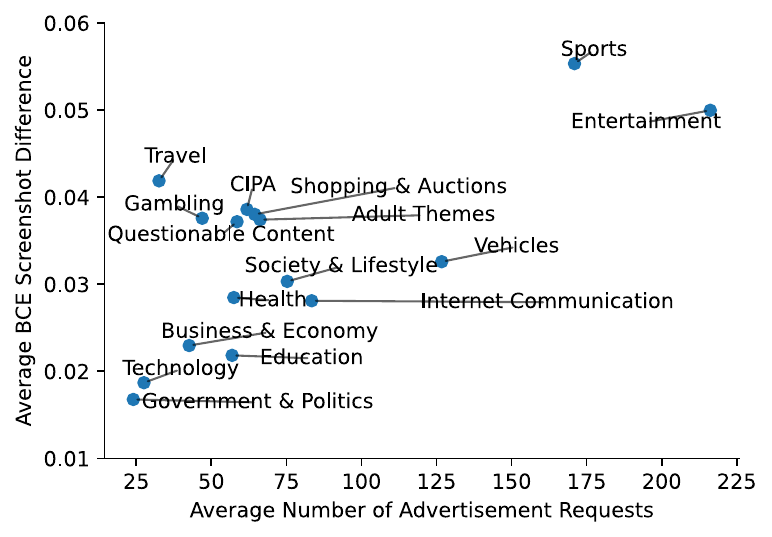}
    \Description{Sports and entertainment have the highest average BCE screenshot difference. Government/politics and technology websites have the lowest average BCE screenshot difference. There is a positive correlation between number of advertisement requests and BCE screenshot difference.}
\end{center}
\caption{\label{fig:easylist_vs_bce} Average number of advertisement requests versus average BCE screenshot difference.}
\end{figure}

We first compare screenshots using Algorithm~\ref{algo:bce_comparison}. The distribution of differences is plotted as a CDF in Figure~\ref{fig:bce_diff}. We find that more than $90\%$ of domains exhibit less than a $10\%$ screenshot difference when cookies are disabled. These results suggest that disabling all third-party cookies does not introduce a substantial visual effect on most websites. Note that all curves in the figure are very similar, indicating that our conclusion is robust across different clickstream lengths used by the crawler.

We also compare the extracted frequency vectors. CDFs of frequency vector DiDs are plotted in Figure~\ref{fig:combined-jaccard-did}. 
The distribution is concentrated and symmetric at $0$. We present the results separated by clickstream length in Appendix~\ref{app:did_depth}. These results suggest that disabling all third-party cookies generally has little effect on the content of a website.

Additionally, since different types of websites may rely on cookies to varying extents, we broke down our analysis by website category. To categorize websites, we used the Cloudflare Domain Intelligence API~\cite{CloudflareDomainIntelligence2024} which has been demonstrated to be reasonably accurate~\cite{ruthWorldWideView2022, ruthTopplingTopLists2022}. After filtering out miscellaneous/uncategorized websites and categories with less than 50 samples, we arrived at a final collection of 16 categories. Table~\ref{tab:category_values} shows the number of websites in each category.

Using the BCE screenshot difference, we found that disabling third-party cookies results in the highest screenshot differences for sports (5.5\%) and entertainment (5.0\%) websites and the lowest screenshot difference for government \& politics (1.7\%) websites. This difference is likely due to the increased use of targeted advertising in sports and entertainment websites. Through manual inspection, we found examples where our crawler saw the same advertisements during the baseline and control group but different advertisements in the experimental group. To more broadly support this claim, we also measured the number of requests made to the advertisement domains in EasyList~\cite{EasyListOverview2024}. As shown by Figure~\ref{fig:easylist_vs_bce}, there is a positive correlation between the average number of advertisement requests and the average BCE screenshot difference across the categorized websites. While removing third-party cookies generally results in minor visual effects, websites that rely heavily on advertisements (such as sports or entertainment websites) often display larger visual changes.
\section{Discussion}
\label{sec:discussion}

In our study, we modeled user web browsing behavior as clickstreams. This approach allowed for the measurement of the inner pages of a website, a consideration that previous studies~\cite{aqeelLandingInternalWeb2020} have emphasized. This approach also facilitated the creation of multiple crawl groups that differed only in their experimental condition (i.e., whether cookies were enabled or disabled). However, there are two important limitations of this clickstream model:
\begin{enumerate}
    \item \textbf{Dynamic content:} Because many web pages are dynamic, generated clickstreams may not be stable across time. Therefore, our dataset is biased towards static inner pages which may cause us to underestimate the impact cookies have on website appearance, as we are less likely to explore the more dynamic areas of a website.
    
    \item \textbf{Random traversal:} To generate a clickstream, we repeatedly sample uniformly at random from the current set of clickable elements. However, random traversal does not capture the intricacies of real user behavior such as typing in search inputs or signing in to a website. Therefore, we may underestimate the effect of disabling third-party cookies.
    To model \textit{purposeful} web browsing, we must use a more targeted clickstream generation method. Unfortunately, generating targeted clickstreams is a difficult problem. White et al. has shown that there is high variability in the behavior of search engine users in their issued query, clicked result, and post-query browsing~\cite{whiteInvestigatingBehavioralVariability2007, whiteInvestigatingQueryingBrowsing2007}. This implies that there is no "average" user that can be modeled. Because of these difficulties, we leave the task of generating realistic clickstreams as future work.
\end{enumerate}

We run all measurements with Firefox. It is unclear whether our conclusions still hold with different browser vendors, such as Chrome or Safari. We leave the investigation on other browsers as future work.

Our results suggest that disabling third-party cookies does not substantially affect the rendering behavior of most websites. One typical effect of cookieless browsing is that users will see different advertisements on certain types of websites, such as those related to sports and entertainment. This provides validation to the industry-wide shift towards cookieless browsing. 
Advertisers will either need to adopt cookie substitutions (such as the APIs proposed in Google's Privacy Sandbox) or move back to non-targeted advertising~\cite{FutureTargetedAdvertising2024}. 

As the majority of third-party cookies are \textit{tracking} cookies~\cite{cahnEmpiricalStudyWeb2016}, disabling all third-party cookies will certainly improve user privacy. However, domains can still track users across websites using only \textit{first-party} cookies, through techniques such as cookie respawning~\cite{fouadDidDeleteMy2021} or first-party cookie leakage~\cite{chenCookieSwapParty2021}. Thus, a natural extension of the present study is to examine how disabling \textit{all} web cookies affects website behavior.
\section{Conclusion}
\label{sec:conclusion}

In this study, we develop a framework to determine whether a given website requires the use of third-party cookies. Specifically, we devise different groups which crawl the same clickstream with varying experimental conditions. We implement this framework in a Selenium-based web crawler and deploy it across the top 10,000 Tranco domains. By comparing extracted screenshots, text, images, and links, we conclude that third-party cookies do not have an observable visual impact on the vast majority of websites.

\begin{acks}
    We sincerely thank the anonymous shepherd and reviewers for their valuable feedback. This study is supported in part by the Duke CS+ summer research program and NSF award CNS-2225448.
\end{acks}


\bibliographystyle{ACM-Reference-Format}
\balance
\bibliography{bib}


\begin{thebibliography}{50}


\ifx \showCODEN    \undefined \def \showCODEN     #1{\unskip}     \fi
\ifx \showDOI      \undefined \def \showDOI       #1{#1}\fi
\ifx \showISBNx    \undefined \def \showISBNx     #1{\unskip}     \fi
\ifx \showISBNxiii \undefined \def \showISBNxiii  #1{\unskip}     \fi
\ifx \showISSN     \undefined \def \showISSN      #1{\unskip}     \fi
\ifx \showLCCN     \undefined \def \showLCCN      #1{\unskip}     \fi
\ifx \shownote     \undefined \def \shownote      #1{#1}          \fi
\ifx \showarticletitle \undefined \def \showarticletitle #1{#1}   \fi
\ifx \showURL      \undefined \def \showURL       {\relax}        \fi
\providecommand\bibfield[2]{#2}
\providecommand\bibinfo[2]{#2}
\providecommand\natexlab[1]{#1}
\providecommand\showeprint[2][]{arXiv:#2}

\bibitem[Dir(2002)]%
        {Directive2002582002}
 \bibinfo{year}{2002}\natexlab{}.
\newblock \bibinfo{title}{{Directive on Privacy and Electronic Communications}}.
\newblock
\newblock
\urldef\tempurl%
\url{https://data.europa.eu/eli/dir/2002/58/oj/eng}
\showURL{%
\tempurl}


\bibitem[ICC(2012)]%
        {ICCUKCookie2012}
 \bibinfo{year}{2012}\natexlab{}.
\newblock \bibinfo{title}{{{ICC UK Cookie Guide}}}.
\newblock \bibinfo{howpublished}{\url{https://www.cookielaw.org/wp-content/uploads/2019/12/icc_uk_cookiesguide_revnov.pdf}}.
\newblock


\bibitem[GDP(2016)]%
        {GDPR2016a}
 \bibinfo{year}{2016}\natexlab{}.
\newblock \bibinfo{title}{{General Data Protection Regulation (GDPR)}}.
\newblock
\newblock
\urldef\tempurl%
\url{https://data.europa.eu/eli/reg/2016/679/oj}
\showURL{%
\tempurl}


\bibitem[Lat(2018)]%
        {LatestFirefoxRolls2018}
 \bibinfo{year}{2018}\natexlab{}.
\newblock \bibinfo{title}{Latest {Firefox} {Rolls} {Out} {Enhanced} {Tracking} {Protection}}.
\newblock
\newblock
\urldef\tempurl%
\url{https://blog.mozilla.org/en/products/firefox/latest-firefox-rolls-out-enhanced-tracking-protection/}
\showURL{%
\tempurl}


\bibitem[Ful(2020)]%
        {FullThirdPartyCookie2020}
 \bibinfo{year}{2020}\natexlab{}.
\newblock \bibinfo{title}{Full {Third}-{Party} {Cookie} {Blocking} and {More}}.
\newblock
\newblock
\urldef\tempurl%
\url{https://webkit.org/blog/10218/full-third-party-cookie-blocking-and-more/}
\showURL{%
\tempurl}


\bibitem[Cal(2024)]%
        {CaliforniaConsumerPrivacy2018}
 \bibinfo{year}{2024}\natexlab{}.
\newblock \bibinfo{title}{California {{Consumer Privacy Act}} ({{CCPA}})}.
\newblock
\newblock
\urldef\tempurl%
\url{https://oag.ca.gov/privacy/ccpa}
\showURL{%
\tempurl}


\bibitem[CIP(2024)]%
        {CIPA}
 \bibinfo{year}{2024}\natexlab{}.
\newblock \bibinfo{title}{{Children's Internet Protection Act (CIPA)}}.
\newblock
\newblock
\urldef\tempurl%
\url{https://www.fcc.gov/consumers/guides/childrens-internet-protection-act}
\showURL{%
\tempurl}


\bibitem[Clo(2024a)]%
        {CloudflareDomainIntelligence2024}
 \bibinfo{year}{2024}\natexlab{a}.
\newblock \bibinfo{title}{Cloudflare {Domain} {Intelligence} {API}}.
\newblock
\newblock
\urldef\tempurl%
\url{https://developers.cloudflare.com/api/operations/domain-intelligence-get-multiple-domain-details}
\showURL{%
\tempurl}


\bibitem[Coo(2024)]%
        {CookieScript}
 \bibinfo{year}{2024}\natexlab{}.
\newblock \bibinfo{title}{Cookie-{{Script}}}.
\newblock
\newblock
\urldef\tempurl%
\url{https://cookie-script.com/}
\showURL{%
\tempurl}


\bibitem[Dis(2024)]%
        {Disconnect2024}
 \bibinfo{year}{2024}\natexlab{}.
\newblock \bibinfo{title}{Disconnect}.
\newblock
\newblock
\urldef\tempurl%
\url{https://disconnect.me/trackerprotection}
\showURL{%
\tempurl}


\bibitem[Clo(2024b)]%
        {CloudflareDocs}
 \bibinfo{year}{2024}\natexlab{b}.
\newblock \bibinfo{title}{{Domain Categories}}.
\newblock
\newblock
\urldef\tempurl%
\url{https://developers.cloudflare.com/cloudflare-one/policies/gateway/domain-categories/}
\showURL{%
\tempurl}


\bibitem[Eas(2024)]%
        {EasyListOverview2024}
 \bibinfo{year}{2024}\natexlab{}.
\newblock \bibinfo{title}{{EasyList}}.
\newblock
\newblock
\urldef\tempurl%
\url{https://easylist.to/}
\showURL{%
\tempurl}


\bibitem[Fut(2024)]%
        {FutureTargetedAdvertising2024}
 \bibinfo{year}{2024}\natexlab{}.
\newblock \bibinfo{title}{The {Future} of {Targeted} {Advertising} in a {Cookie}-less {World}}.
\newblock
\newblock
\urldef\tempurl%
\url{https://insight.kellogg.northwestern.edu/article/online-consumer-privacy-advertising}
\showURL{%
\tempurl}


\bibitem[tea(2024)]%
        {teamGetReadyCookieless2024}
 \bibinfo{year}{2024}\natexlab{}.
\newblock \bibinfo{title}{{Get Ready for the Cookieless Future and the End of Third-Party Cookies}}.
\newblock
\newblock
\urldef\tempurl%
\url{https://business.adobe.com/blog/the-latest/prepare-cookieless-future}
\showURL{%
\tempurl}


\bibitem[WIR(2024)]%
        {WIREDBrandLab2024}
 \bibinfo{year}{2024}\natexlab{}.
\newblock \bibinfo{title}{{How} {Companies} {Can} {Thrive} in a {Cookieless} {Future}}.
\newblock
\newblock
\urldef\tempurl%
\url{https://www.wired.com/sponsored/story/how-companies-can-thrive-in-a-cookieless-future/}
\showURL{%
\tempurl}


\bibitem[Pre(2024)]%
        {PreparePhasingOut}
 \bibinfo{year}{2024}\natexlab{}.
\newblock \bibinfo{title}{{Prepare for Phasing Out Third-Party Cookies}}.
\newblock
\newblock
\urldef\tempurl%
\url{https://developers.google.com/privacy-sandbox/3pcd}
\showURL{%
\tempurl}


\bibitem[Sel(2024)]%
        {Selenium}
 \bibinfo{year}{2024}\natexlab{}.
\newblock \bibinfo{title}{{Selenium WebDriver}}.
\newblock
\newblock
\urldef\tempurl%
\url{https://www.selenium.dev/documentation/webdriver/}
\showURL{%
\tempurl}


\bibitem[Anderson et~al\mbox{.}(2007)]%
        {andersonSpamscatterCharacterizingInternet2007}
\bibfield{author}{\bibinfo{person}{David~S. Anderson}, \bibinfo{person}{Chris Fleizach}, \bibinfo{person}{Stefan Savage}, {and} \bibinfo{person}{Geoffrey~M. Voelker}.} \bibinfo{year}{2007}\natexlab{}.
\newblock \showarticletitle{{Spamscatter: Characterizing Internet Scam Hosting Infrastructure}}. In \bibinfo{booktitle}{\emph{USENIX Security}}. USENIX.
\newblock


\bibitem[Aqeel et~al\mbox{.}(2020)]%
        {aqeelLandingInternalWeb2020}
\bibfield{author}{\bibinfo{person}{Waqar Aqeel}, \bibinfo{person}{Balakrishnan Chandrasekaran}, \bibinfo{person}{Anja Feldmann}, {and} \bibinfo{person}{Bruce~M. Maggs}.} \bibinfo{year}{2020}\natexlab{}.
\newblock \showarticletitle{{On Landing and Internal Web Pages: The Strange Case of Jekyll and Hyde in Web Performance Measurement}}. In \bibinfo{booktitle}{\emph{Proceedings of IMC}}. ACM, \bibinfo{pages}{680–695}.
\newblock


\bibitem[Bollinger et~al\mbox{.}(2022)]%
        {bollingerAutomatingCookieConsent2022}
\bibfield{author}{\bibinfo{person}{Dino Bollinger}, \bibinfo{person}{Karel Kubicek}, \bibinfo{person}{Carlos Cotrini}, {and} \bibinfo{person}{David Basin}.} \bibinfo{year}{2022}\natexlab{}.
\newblock \showarticletitle{{Automating Cookie Consent and {GDPR} Violation Detection}}. In \bibinfo{booktitle}{\emph{USENIX Security}}. USENIX, \bibinfo{pages}{2893--2910}.
\newblock


\bibitem[{Cahn, Aaron and Alfeld, Scott and Barford, Paul and Muthukrishnan, S.}(2016)]%
        {cahnEmpiricalStudyWeb2016}
\bibfield{author}{\bibinfo{person}{{Cahn, Aaron and Alfeld, Scott and Barford, Paul and Muthukrishnan, S.}}} \bibinfo{year}{2016}\natexlab{}.
\newblock \showarticletitle{{An Empirical Study of Web Cookies}}. In \bibinfo{booktitle}{\emph{Proceedings of WWW}}. ACM, \bibinfo{pages}{891–901}.
\newblock


\bibitem[Chen et~al\mbox{.}(2021)]%
        {chenCookieSwapParty2021}
\bibfield{author}{\bibinfo{person}{Quan Chen}, \bibinfo{person}{Panagiotis Ilia}, \bibinfo{person}{Michalis Polychronakis}, {and} \bibinfo{person}{Alexandros Kapravelos}.} \bibinfo{year}{2021}\natexlab{}.
\newblock \showarticletitle{{Cookie Swap Party: Abusing First-Party Cookies for Web Tracking}}. In \bibinfo{booktitle}{\emph{Proceedings of WWW}}. ACM, \bibinfo{pages}{2117--2129}.
\newblock


\bibitem[Degeling et~al\mbox{.}(2019)]%
        {degelingWeValueYour2019}
\bibfield{author}{\bibinfo{person}{Martin Degeling}, \bibinfo{person}{Christine Utz}, \bibinfo{person}{Christopher Lentzsch}, \bibinfo{person}{Henry Hosseini}, \bibinfo{person}{Florian Schaub}, {and} \bibinfo{person}{Thorsten Holz}.} \bibinfo{year}{2019}\natexlab{}.
\newblock \showarticletitle{We {Value} {Your} {Privacy} ... {Now} {Take} {Some} {Cookies}: {Measuring} the {GDPR}'s {Impact} on {Web} {Privacy}}. In \bibinfo{booktitle}{\emph{Proceedings of NDSS}}. Internet Society, \bibinfo{pages}{345--346}.
\newblock


\bibitem[Englehardt and Narayanan(2016)]%
        {CCS16-OnlineTracking}
\bibfield{author}{\bibinfo{person}{Steven Englehardt} {and} \bibinfo{person}{Arvind Narayanan}.} \bibinfo{year}{2016}\natexlab{}.
\newblock \showarticletitle{{Online Tracking: A 1-Million-Site Measurement and Analysis}}. In \bibinfo{booktitle}{\emph{Proceedings of CCS}}. ACM, \bibinfo{pages}{1388--1401}.
\newblock


\bibitem[Englehardt et~al\mbox{.}(2015)]%
        {englehardtCookiesThatGive2015}
\bibfield{author}{\bibinfo{person}{Steven Englehardt}, \bibinfo{person}{Dillon Reisman}, \bibinfo{person}{Christian Eubank}, \bibinfo{person}{Peter Zimmerman}, \bibinfo{person}{Jonathan Mayer}, \bibinfo{person}{Arvind Narayanan}, {and} \bibinfo{person}{Edward~W. Felten}.} \bibinfo{year}{2015}\natexlab{}.
\newblock \showarticletitle{{Cookies That Give You Away: The Surveillance Implications of Web Tracking}}. In \bibinfo{booktitle}{\emph{Proceedings of WWW}}. ACM, \bibinfo{pages}{289–299}.
\newblock


\bibitem[Fouad et~al\mbox{.}(2020)]%
        {fouadComplianceCookiePurposes2020}
\bibfield{author}{\bibinfo{person}{Imane Fouad}, \bibinfo{person}{Cristiana Santos}, \bibinfo{person}{Feras Al~Kassar}, \bibinfo{person}{Nataliia Bielova}, {and} \bibinfo{person}{Stefano Calzavara}.} \bibinfo{year}{2020}\natexlab{}.
\newblock \showarticletitle{On {Compliance} of {Cookie} {Purposes} with the {Purpose} {Specification} {Principle}}. In \bibinfo{booktitle}{\emph{EuroS\&PW 2020}}. {IEEE}, \bibinfo{pages}{326--333}.
\newblock


\bibitem[Fouad et~al\mbox{.}(2021)]%
        {fouadDidDeleteMy2021}
\bibfield{author}{\bibinfo{person}{Imane Fouad}, \bibinfo{person}{Cristiana Santos}, \bibinfo{person}{Arnaud Legout}, {and} \bibinfo{person}{Nataliia Bielova}.} \bibinfo{year}{2021}\natexlab{}.
\newblock \bibinfo{title}{{Did I Delete My Cookies? Cookies Respawning with Browser Fingerprinting}}.
\newblock
\newblock
\showeprint[arxiv]{2105.04381}~[cs.CR]


\bibitem[Fridrich(2024)]%
        {fridrichFczbkkCssselectorgeneratorbenchmarkBenchmark2024}
\bibfield{author}{\bibinfo{person}{Riki Fridrich}.} \bibinfo{year}{2024}\natexlab{}.
\newblock \bibinfo{title}{{CSS Selector Generator Benchmark}}.
\newblock
\newblock
\urldef\tempurl%
\url{https://github.com/fczbkk/css-selector-generator-benchmark}
\showURL{%
\tempurl}


\bibitem[Habib et~al\mbox{.}(2022)]%
        {habibOkayWhateverEvaluation2022}
\bibfield{author}{\bibinfo{person}{Hana Habib}, \bibinfo{person}{Megan Li}, \bibinfo{person}{Ellie Young}, {and} \bibinfo{person}{Lorrie Cranor}.} \bibinfo{year}{2022}\natexlab{}.
\newblock \showarticletitle{{“Okay, whatever”: An Evaluation of Cookie Consent Interfaces}}. In \bibinfo{booktitle}{\emph{Proceedings of CHI}}. ACM, \bibinfo{pages}{1--27}.
\newblock


\bibitem[Hoffman et~al\mbox{.}(2019)]%
        {hoffmanDNSTerminology2019}
\bibfield{author}{\bibinfo{person}{Paul~E. Hoffman}, \bibinfo{person}{Anew Sullivan}, {and} \bibinfo{person}{Kazunori Fujiwara}.} \bibinfo{year}{2019}\natexlab{}.
\newblock \bibinfo{booktitle}{\emph{{DNS} {Terminology}}}.
\newblock \bibinfo{type}{Request for {Comments}} RFC 8499. \bibinfo{institution}{Internet Engineering Task Force}.
\newblock
\urldef\tempurl%
\url{https://datatracker.ietf.org/doc/rfc8499}
\showURL{%
\tempurl}


\bibitem[Hu et~al\mbox{.}(2021)]%
        {huCCCCCorrallingCookies2021}
\bibfield{author}{\bibinfo{person}{Xuehui Hu}, \bibinfo{person}{Nishanth Sastry}, {and} \bibinfo{person}{Mainack Mondal}.} \bibinfo{year}{2021}\natexlab{}.
\newblock \showarticletitle{{CCCC: Corralling Cookies into Categories with CookieMonster}}. In \bibinfo{booktitle}{\emph{Proceedings of WebSci}}. ACM, \bibinfo{pages}{234–242}.
\newblock


\bibitem[Kampanos and Shahandashti(2021)]%
        {kampanosAcceptAllLandscape2021}
\bibfield{author}{\bibinfo{person}{Georgios Kampanos} {and} \bibinfo{person}{Siamak~F. Shahandashti}.} \bibinfo{year}{2021}\natexlab{}.
\newblock \showarticletitle{{Accept All: The Landscape of Cookie Banners in Greece and the UK}}. In \bibinfo{booktitle}{\emph{ICT Systems Security and Privacy Protection}}. Springer, \bibinfo{pages}{213--227}.
\newblock


\bibitem[Kelman et~al\mbox{.}(2007)]%
        {kelmanKeypointDescriptorsMatching2007}
\bibfield{author}{\bibinfo{person}{Avi Kelman}, \bibinfo{person}{Michal Sofka}, {and} \bibinfo{person}{Charles~V. Stewart}.} \bibinfo{year}{2007}\natexlab{}.
\newblock \showarticletitle{{Keypoint Descriptors for Matching Across Multiple Image Modalities and Non-linear Intensity Variations}}. In \bibinfo{booktitle}{\emph{Proceedings of CVPR}}. IEEE, \bibinfo{pages}{1--7}.
\newblock


\bibitem[Khandelwal et~al\mbox{.}(2023)]%
        {khandelwalAutomatedCookieNotice2023}
\bibfield{author}{\bibinfo{person}{Rishabh Khandelwal}, \bibinfo{person}{Asmit Nayak}, \bibinfo{person}{Hamza Harkous}, {and} \bibinfo{person}{Kassem Fawaz}.} \bibinfo{year}{2023}\natexlab{}.
\newblock \showarticletitle{{Automated Cookie Notice Analysis and Enforcement}}. In \bibinfo{booktitle}{\emph{USENIX Security}}. USENIX, \bibinfo{pages}{1109--1126}.
\newblock


\bibitem[{Le Pochat} et~al\mbox{.}(2019)]%
        {LePochat2019}
\bibfield{author}{\bibinfo{person}{Victor {Le Pochat}}, \bibinfo{person}{Tom {Van Goethem}}, \bibinfo{person}{Samaneh Tajalizadehkhoob}, \bibinfo{person}{Maciej Korczy\'{n}ski}, {and} \bibinfo{person}{Wouter Joosen}.} \bibinfo{year}{2019}\natexlab{}.
\newblock \showarticletitle{{Tranco: A Research-Oriented Top Sites Ranking Hardened Against Manipulation}}. In \bibinfo{booktitle}{\emph{Proceedings of NDSS}}.
\newblock


\bibitem[Lehmann et~al\mbox{.}(2013)]%
        {lehmannOnlineMultitaskingUser2013a}
\bibfield{author}{\bibinfo{person}{Janette Lehmann}, \bibinfo{person}{Mounia Lalmas}, \bibinfo{person}{Georges Dupret}, {and} \bibinfo{person}{Ricardo Baeza-Yates}.} \bibinfo{year}{2013}\natexlab{}.
\newblock \showarticletitle{{Online Multitasking and User Engagement}}. In \bibinfo{booktitle}{\emph{Proceedings of CIKM}}. ACM, \bibinfo{pages}{519–528}.
\newblock


\bibitem[Lerner et~al\mbox{.}(2016)]%
        {Security16-WebTracking}
\bibfield{author}{\bibinfo{person}{Ada Lerner}, \bibinfo{person}{Anna~Kornfeld Simpson}, \bibinfo{person}{Tadayoshi Kohno}, {and} \bibinfo{person}{Franziska Roesner}.} \bibinfo{year}{2016}\natexlab{}.
\newblock \showarticletitle{{Internet Jones and the Raiders of the Lost Trackers: An Archaeological Study of Web Tracking from 1996 to 2016}}. In \bibinfo{booktitle}{\emph{USENIX Security}}. USENIX.
\newblock


\bibitem[Li et~al\mbox{.}(2015)]%
        {PAM15-TrackAdvisor}
\bibfield{author}{\bibinfo{person}{Tai-Ching Li}, \bibinfo{person}{Huy Hang}, \bibinfo{person}{Michalis Faloutsos}, {and} \bibinfo{person}{Petros Efstathopoulos}.} \bibinfo{year}{2015}\natexlab{}.
\newblock \showarticletitle{{TrackAdvisor: Taking Back Browsing Privacy from Third-Party Trackers}}. In \bibinfo{booktitle}{\emph{Proceedings of PAM}}. Springer, \bibinfo{pages}{277--289}.
\newblock


\bibitem[Lourenco et~al\mbox{.}(2012)]%
        {lourencoSRDSIFTKeypointDetection2012}
\bibfield{author}{\bibinfo{person}{Miguel Lourenco}, \bibinfo{person}{Joao~P. Barreto}, {and} \bibinfo{person}{Francisco Vasconcelos}.} \bibinfo{year}{2012}\natexlab{}.
\newblock \showarticletitle{{sRD-SIFT: Keypoint Detection and Matching in Images With Radial Distortion}}.
\newblock \bibinfo{journal}{\emph{IEEE Transactions on Robotics}} \bibinfo{volume}{28}, \bibinfo{number}{3} (\bibinfo{year}{2012}), \bibinfo{pages}{752--760}.
\newblock


\bibitem[Medvedev(2024)]%
        {medvedevAntonmedvFinder2024}
\bibfield{author}{\bibinfo{person}{Anton Medvedev}.} \bibinfo{year}{2024}\natexlab{}.
\newblock \bibinfo{title}{finder}.
\newblock
\newblock
\urldef\tempurl%
\url{https://github.com/antonmedv/finder}
\showURL{%
\tempurl}


\bibitem[Pal et~al\mbox{.}(2023)]%
        {ECommerce23-Clickstream}
\bibfield{author}{\bibinfo{person}{Gautam Pal}, \bibinfo{person}{Katie Atkinson}, {and} \bibinfo{person}{Gangmin Li}.} \bibinfo{year}{2023}\natexlab{}.
\newblock \showarticletitle{{Real-Time User Clickstream Behavior Analysis based on Apache Storm Streaming}}.
\newblock \bibinfo{journal}{\emph{Electronic Commerce Research}} \bibinfo{volume}{23}, \bibinfo{number}{3} (\bibinfo{year}{2023}), \bibinfo{pages}{1829--1859}.
\newblock


\bibitem[Papadopoulos et~al\mbox{.}(2019)]%
        {papadopoulosCookieSynchronizationEverything2019}
\bibfield{author}{\bibinfo{person}{Panagiotis Papadopoulos}, \bibinfo{person}{Nicolas Kourtellis}, {and} \bibinfo{person}{Evangelos Markatos}.} \bibinfo{year}{2019}\natexlab{}.
\newblock \showarticletitle{{Cookie Synchronization: Everything You Always Wanted to Know But Were Afraid to Ask}}. In \bibinfo{booktitle}{\emph{Proceedings of WWW}}. ACM, \bibinfo{pages}{1432--1442}.
\newblock


\bibitem[Rasaii et~al\mbox{.}(2023)]%
        {rasaiiExploringCookieverseMultiPerspective2023}
\bibfield{author}{\bibinfo{person}{Ali Rasaii}, \bibinfo{person}{Shivani Singh}, \bibinfo{person}{Devashish Gosain}, {and} \bibinfo{person}{Oliver Gasser}.} \bibinfo{year}{2023}\natexlab{}.
\newblock \showarticletitle{Exploring the {Cookieverse}: {A} {Multi}-{Perspective} {Analysis} of {Web} {Cookies}}.
\newblock In \bibinfo{booktitle}{\emph{Proceedings of PAM}}. \bibinfo{publisher}{Springer}, \bibinfo{pages}{623--651}.
\newblock
\showISBNx{978-3-031-28485-4 978-3-031-28486-1}


\bibitem[Ruth et~al\mbox{.}(2022a)]%
        {ruthWorldWideView2022}
\bibfield{author}{\bibinfo{person}{Kimberly Ruth}, \bibinfo{person}{Aurore Fass}, \bibinfo{person}{Jonathan Azose}, \bibinfo{person}{Mark Pearson}, \bibinfo{person}{Emma Thomas}, \bibinfo{person}{Caitlin Sadowski}, {and} \bibinfo{person}{Zakir Durumeric}.} \bibinfo{year}{2022}\natexlab{a}.
\newblock \showarticletitle{{A World Wide View of Browsing the World Wide Web}}. In \bibinfo{booktitle}{\emph{Proceedings of IMC}}. ACM, \bibinfo{pages}{317–336}.
\newblock


\bibitem[Ruth et~al\mbox{.}(2022b)]%
        {ruthTopplingTopLists2022}
\bibfield{author}{\bibinfo{person}{Kimberly Ruth}, \bibinfo{person}{Deepak Kumar}, \bibinfo{person}{Brandon Wang}, \bibinfo{person}{Luke Valenta}, {and} \bibinfo{person}{Zakir Durumeric}.} \bibinfo{year}{2022}\natexlab{b}.
\newblock \showarticletitle{{Toppling Top Lists: Evaluating the Accuracy of Popular Website Lists}}. In \bibinfo{booktitle}{\emph{Proceedings of IMC}}. ACM, \bibinfo{pages}{374--387}.
\newblock


\bibitem[Wang et~al\mbox{.}(2004)]%
        {ToIP04-SSIM}
\bibfield{author}{\bibinfo{person}{Zhou Wang}, \bibinfo{person}{Alan~C Bovik}, \bibinfo{person}{Hamid~R Sheikh}, {and} \bibinfo{person}{Eero~P Simoncelli}.} \bibinfo{year}{2004}\natexlab{}.
\newblock \showarticletitle{{Image Quality Assessment: From Error Visibility to Structural Similarity}}.
\newblock \bibinfo{journal}{\emph{IEEE Transactions on Image Processing}} \bibinfo{volume}{13}, \bibinfo{number}{4} (\bibinfo{year}{2004}), \bibinfo{pages}{600--612}.
\newblock


\bibitem[White and Drucker(2007)]%
        {whiteInvestigatingBehavioralVariability2007}
\bibfield{author}{\bibinfo{person}{Ryen~W. White} {and} \bibinfo{person}{Steven~M. Drucker}.} \bibinfo{year}{2007}\natexlab{}.
\newblock \showarticletitle{{Investigating Behavioral Variability in Web Search}}. In \bibinfo{booktitle}{\emph{Proceedings of WWW}}. ACM, \bibinfo{pages}{21--30}.
\newblock


\bibitem[White and Morris(2007)]%
        {whiteInvestigatingQueryingBrowsing2007}
\bibfield{author}{\bibinfo{person}{Ryen~W. White} {and} \bibinfo{person}{Dan Morris}.} \bibinfo{year}{2007}\natexlab{}.
\newblock \showarticletitle{{Investigating the Querying and Browsing Behavior of Advanced Search Engine Users}}. In \bibinfo{booktitle}{\emph{Proceedings of SIGIR}}. ACM, \bibinfo{pages}{255–262}.
\newblock


\bibitem[Xing et~al\mbox{.}(2013)]%
        {xingIncorporatingUserPreferences2013a}
\bibfield{author}{\bibinfo{person}{Qianli Xing}, \bibinfo{person}{Yiqun Liu}, \bibinfo{person}{Jian-Yun Nie}, \bibinfo{person}{Min Zhang}, \bibinfo{person}{Shaoping Ma}, {and} \bibinfo{person}{Kuo Zhang}.} \bibinfo{year}{2013}\natexlab{}.
\newblock \showarticletitle{{Incorporating User Preferences into Click Models}}. In \bibinfo{booktitle}{\emph{Proceedings of CIKM}}. ACM, \bibinfo{pages}{1301–1310}.
\newblock


\bibitem[Zhang et~al\mbox{.}(2011)]%
        {ToIP11-FSIM}
\bibfield{author}{\bibinfo{person}{Lin Zhang}, \bibinfo{person}{Lei Zhang}, \bibinfo{person}{Xuanqin Mou}, {and} \bibinfo{person}{David Zhang}.} \bibinfo{year}{2011}\natexlab{}.
\newblock \showarticletitle{{FSIM: A Feature Similarity Index for Image Quality Assessment}}.
\newblock \bibinfo{journal}{\emph{IEEE Transactions on Image Processing}} \bibinfo{volume}{20}, \bibinfo{number}{8} (\bibinfo{year}{2011}), \bibinfo{pages}{2378--2386}.
\newblock


\end{thebibliography}

\appendix
\section{Appendix}

\subsection{Ethics}
This work does not raise any ethical issues.

\subsection{Frequency Vector DiDs Separated by Clickstream Length}\label{app:did_depth}

\begin{figure}[H]
\begin{center}
    \centering
    \includegraphics[width=0.9\columnwidth]{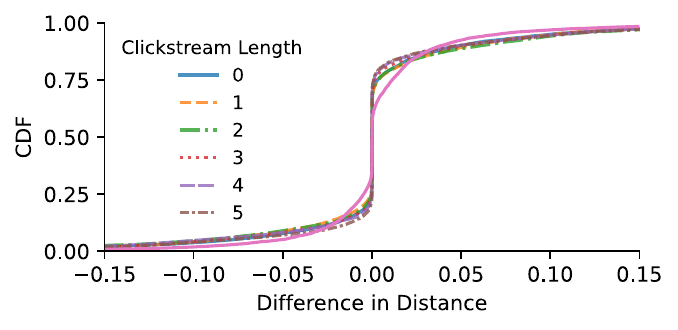}
    \Description{A CDF displaying the difference in distance distribution across image shingles. There is little difference between depths.}
\end{center}
\caption{\textbf{CDF of image shingle DiDs separated by clickstream length.}}
\end{figure}
\begin{figure}[H]
\begin{center}
    \centering
    \includegraphics[width=0.9\columnwidth]{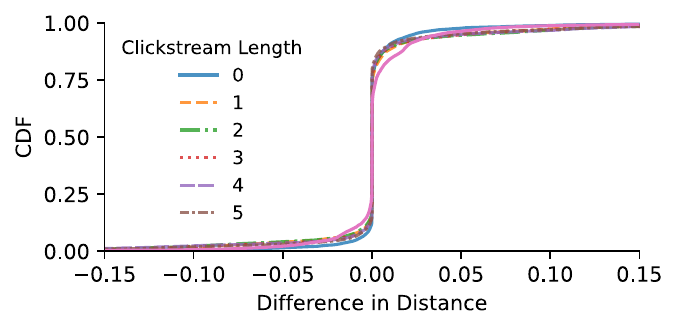}
    \Description{A CDF displaying the difference in distance distribution across inner text. There is little difference between depths.}
\end{center}
\caption{\textbf{CDF of inner text DiDs separated by clickstream length.}}
\end{figure}
\begin{figure}[H]
\begin{center}
    \centering
    \includegraphics[width=0.9\columnwidth]{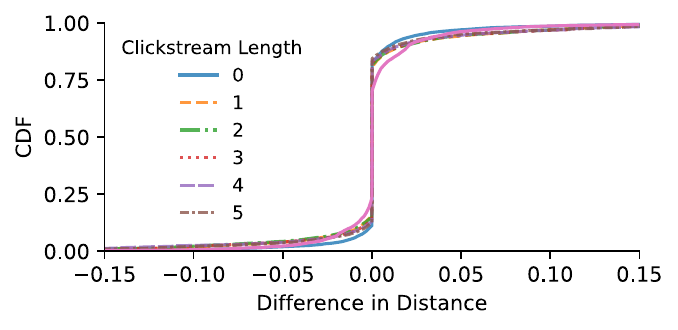}
    \Description{A CDF displaying the difference in distance distribution across images. There is little difference between depths.}
\end{center}
\caption{\textbf{CDF of image DiDs separated by clickstream length.}}
\end{figure}
\begin{figure}[H]
\begin{center}
    \centering
    \includegraphics[width=0.9\columnwidth]{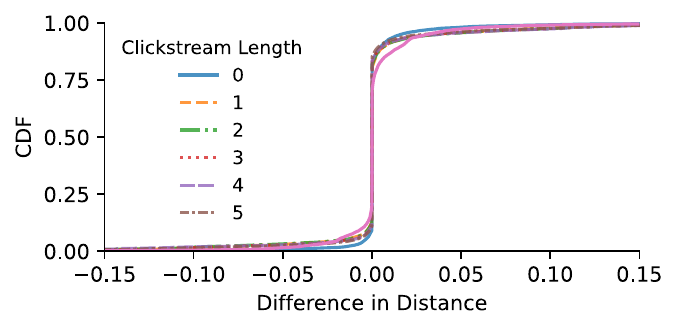}
    \Description{A CDF displaying the difference in distance distribution across links. There is little difference between depths.}
\end{center}
\caption{\textbf{CDF of link DiDs separated by clickstream length.}}
\end{figure}

\subsection{Screenshot Examples}\label{app:example}
The following figures provide some examples of dynamic content and screenshot comparisons.

\begin{figure}[H]
     \centering
     \begin{subfigure}[b]{0.5\textwidth}
         \centering
         \includegraphics[width=\textwidth]{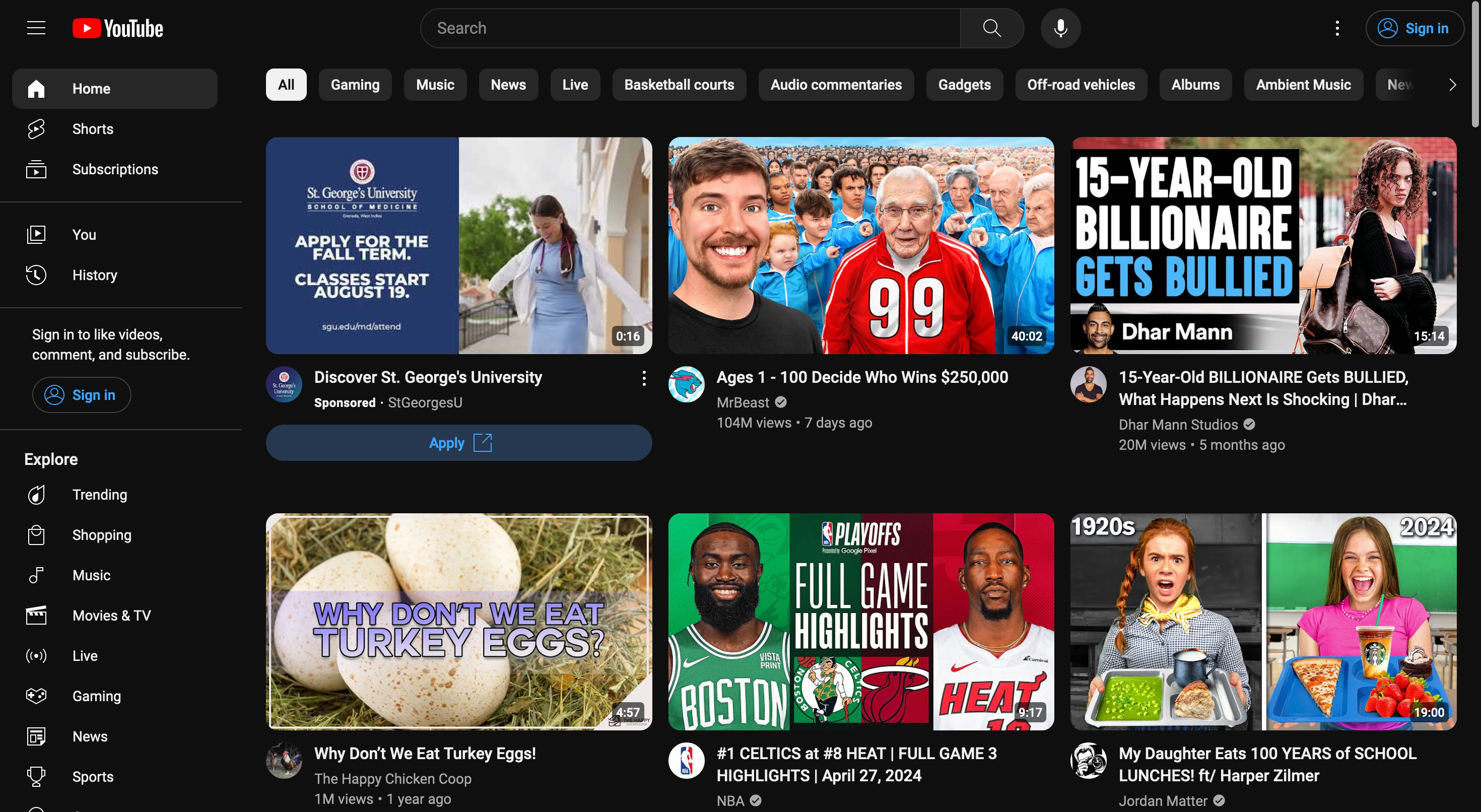}
         \Description{A screenshot of youtube.com.}
     \end{subfigure}
     \hfill
     \begin{subfigure}[b]{0.5\textwidth}
         \centering
         \includegraphics[width=\textwidth]{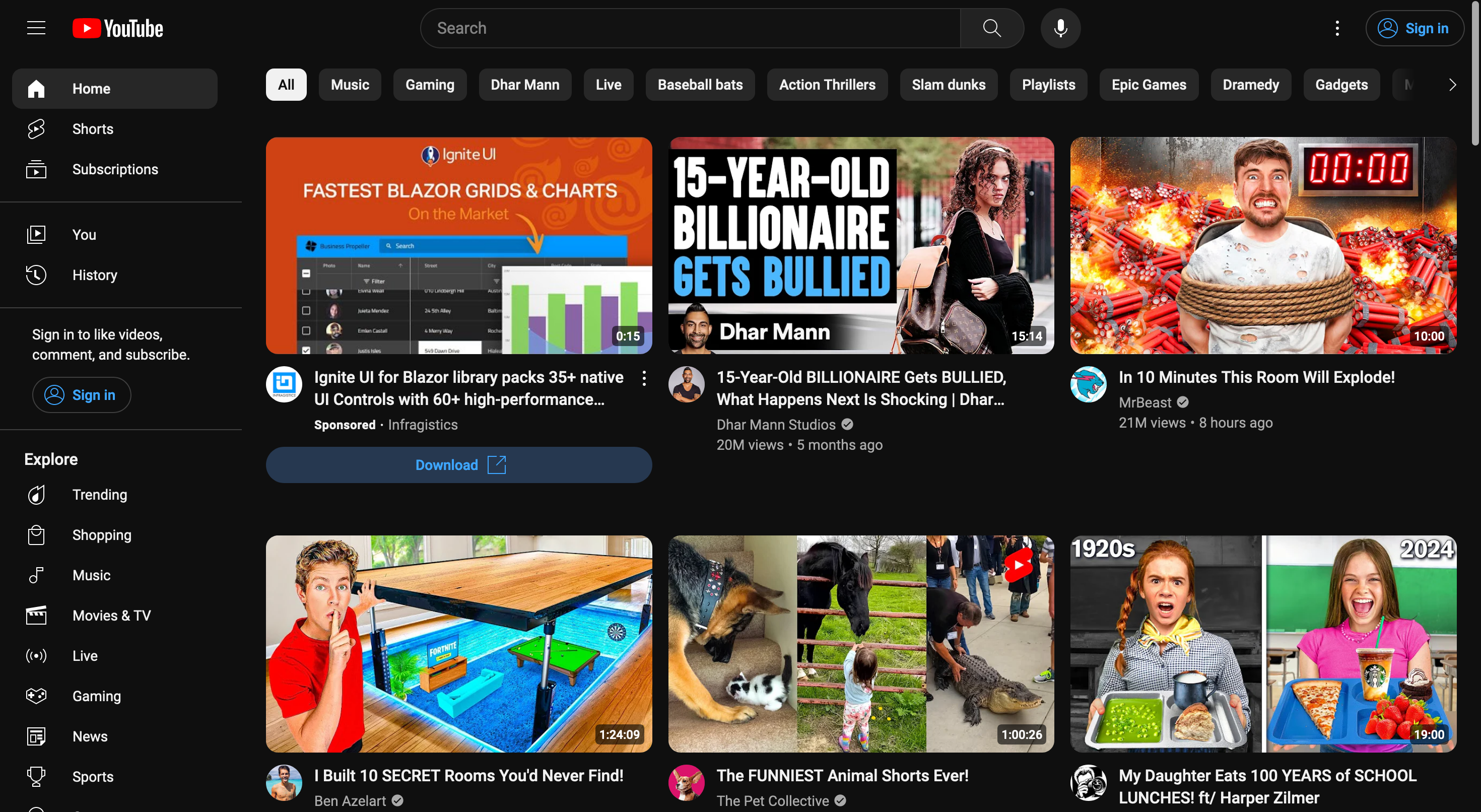}
         \Description{Another screenshot of youtube.com. The recommended videos, search keywords, advertisements have changed since the first screenshot.}
     \end{subfigure}
     \caption{\label{fig:youtube}The website \texttt{youtube.com} displays dynamic content. These two screenshots show how \texttt{youtube.com} changes between page reloads.}
\end{figure}

\balance

\begin{figure}[H]
     \centering
     \begin{subfigure}[b]{0.5\textwidth}
         \centering
         \includegraphics[width=\textwidth]{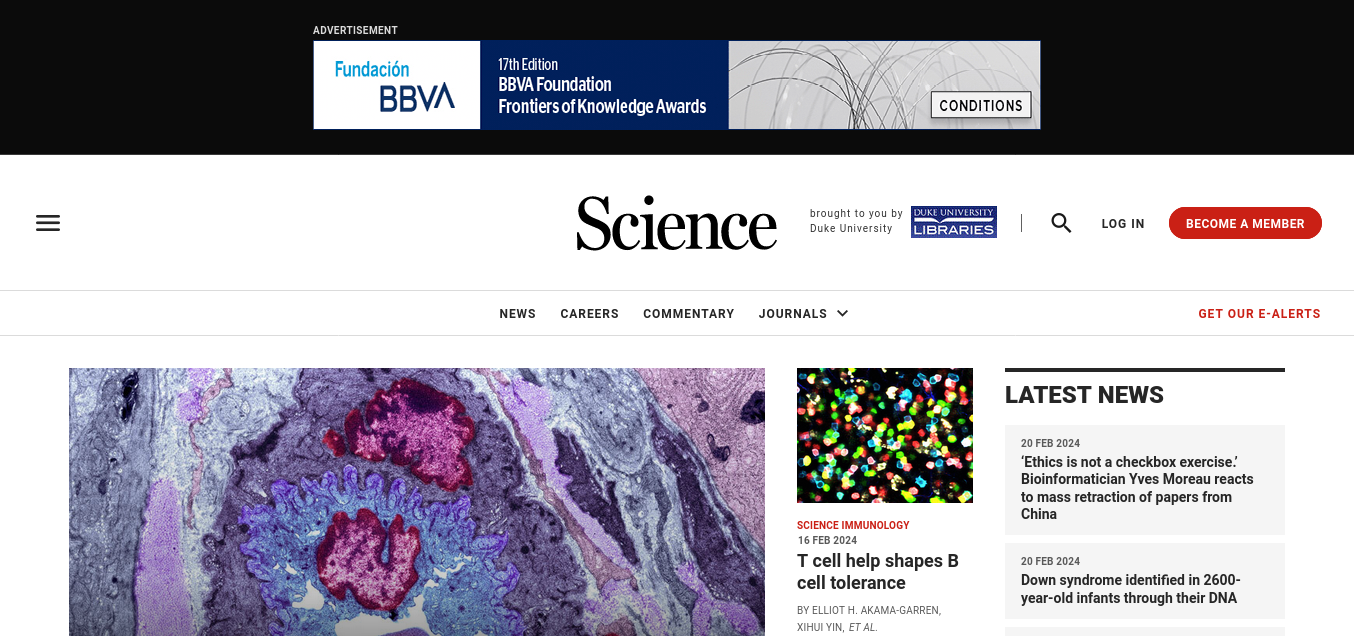}
         \Description{A screenshot of sciencemag.org.}
         \caption{Baseline}
     \end{subfigure}
     \hfill
     \begin{subfigure}[b]{0.5\textwidth}
         \centering
         \includegraphics[width=\textwidth]{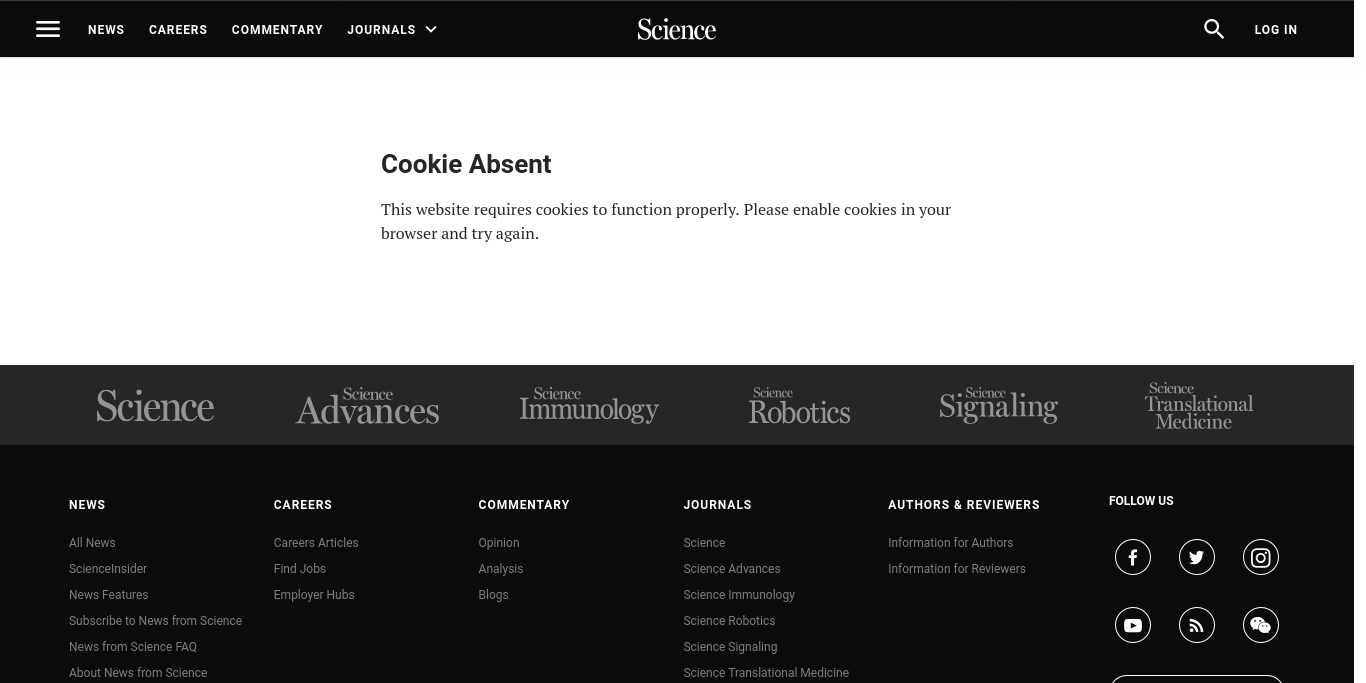}
         \caption{Experimental}
         \Description{Another screenshot of sciencemag.org. An error message about absent cookies is displayed.}
     \end{subfigure}
     \caption{Without cookies, \texttt{sciencemag.org} displays an error message. There is a 0.85 jaccard distance between the image shingles of these two screenshots.}
\end{figure}

\begin{figure}[H]
     \centering
     \begin{subfigure}[b]{0.5\textwidth}
         \centering
         \includegraphics[width=\textwidth]{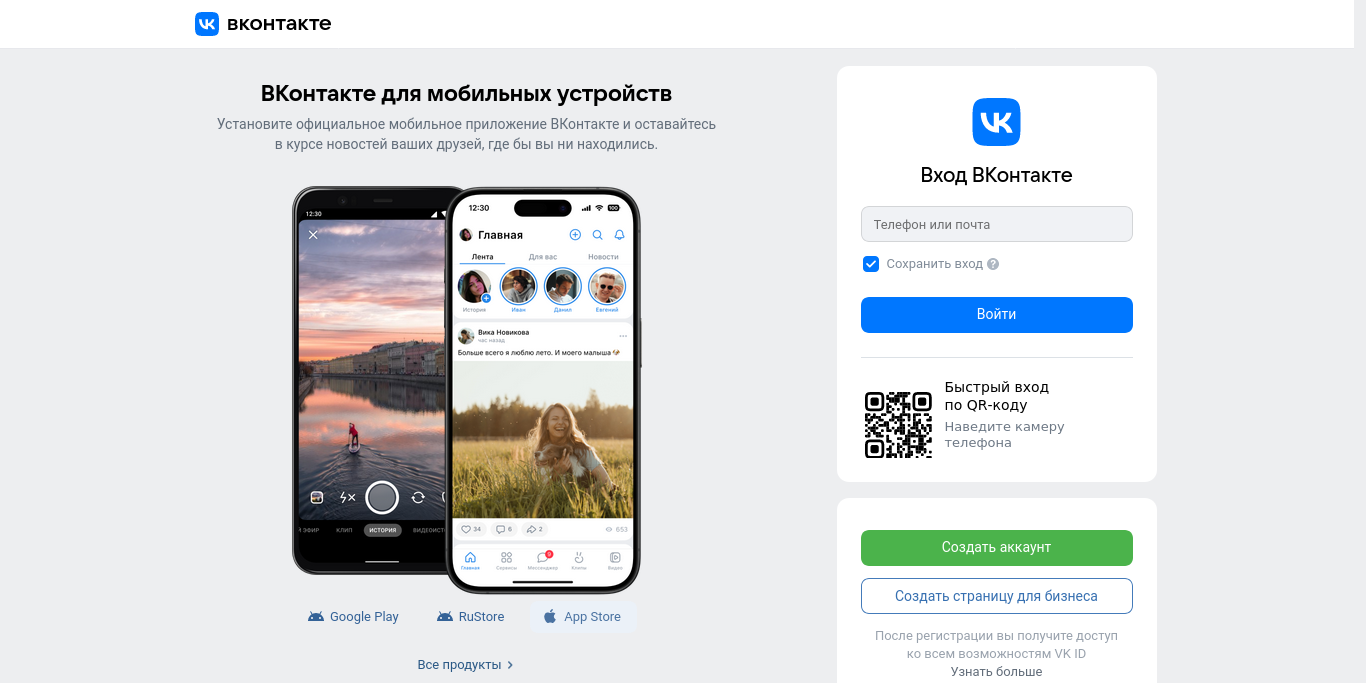}
         \Description{A screenshot of userapi.com. The page is in Russian.}
         \caption{Baseline}
     \end{subfigure}
     \hfill
     \begin{subfigure}[b]{0.5\textwidth}
         \centering
         \includegraphics[width=\textwidth]{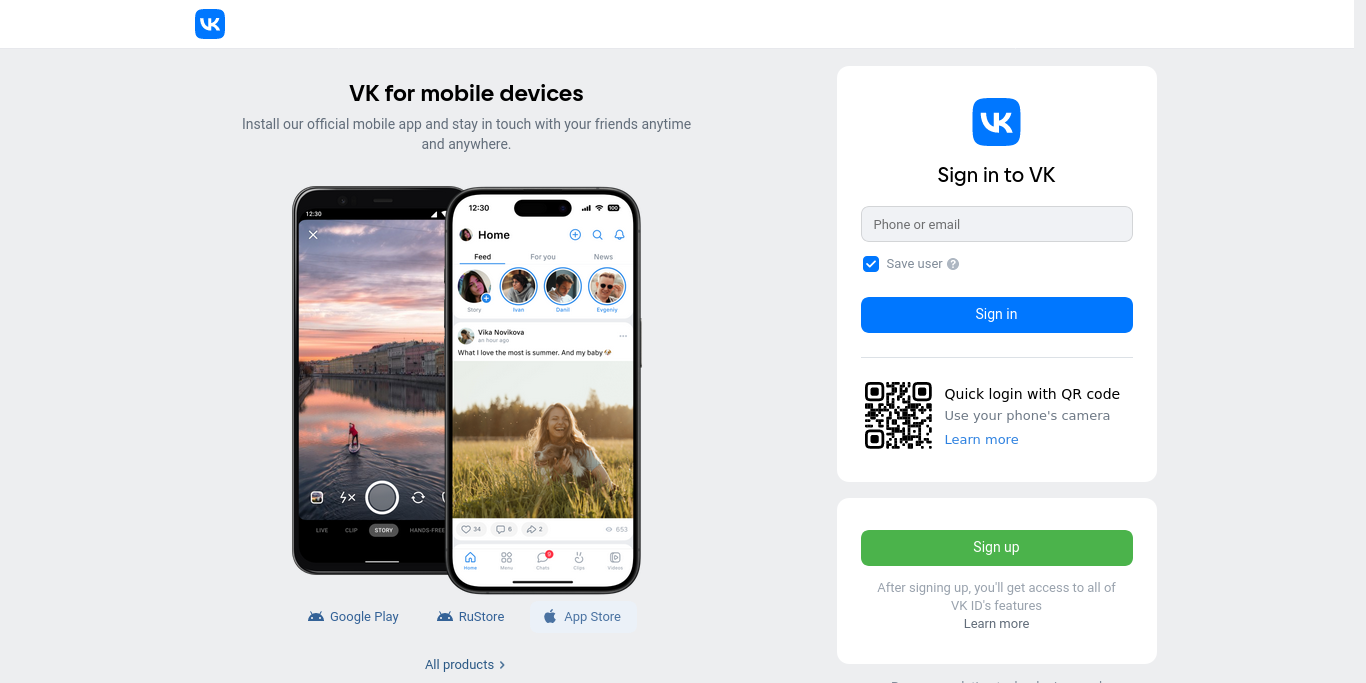}
         \Description{A screenshot of userapi.com. The page is in English.}
         \caption{Experimental}
     \end{subfigure}
     \caption{Without cookies, \texttt{userapi.com} is unable to change languages. There is a 0.35 jaccard distance between the image shingles of these two screenshots.}
\end{figure}

\end{document}